\documentclass[sigconf]{acmart}
\usepackage{amsmath}
\usepackage{amsthm}
\usepackage{mathrsfs}
\usepackage{multirow}
\usepackage{booktabs}
\usepackage{color}
\usepackage{ragged2e}
\usepackage[acronym]{glossaries}
\usepackage{graphicx}
\usepackage{caption}
\usepackage{float}
\usepackage{enumitem}
\usepackage{subcaption}
\usepackage[normalem]{ulem}
\useunder{\uline}{\ul}{}
\makeglossaries
\newacronym{DEISI}{DEISI-GNN}{a GNN-based SBR model that discreetly exploits inter-session information}
\newacronym{DRL}{DRL}{Disentangled Representation Learning}
\newacronym{CL}{CL}{Contrastive learning}
\newacronym{SBR}{SBR}{Session-based Recommendation}
\newacronym{GNN}{GNN}{Graph Neural Network}

\newcommand{\bm}[1]{\mbox{\boldmath{$#1$}}}
\newcommand{\todo}[1]{\textcolor[rgb]{0.0, 0.0, 0.0}{#1}}

\AtBeginDocument{%
  \providecommand\BibTeX{{%
    \normalfont B\kern-0.5em{\scshape i\kern-0.25em b}\kern-0.8em\TeX}}}

\setcopyright{acmcopyright}
\copyrightyear{2018}
\acmYear{2018}
\acmDOI{XXXXXXX.XXXXXXX}

\acmConference[Conference acronym 'XX]{Make sure to enter the correct
  conference title from your rights confirmation emai}{June 03--05,
  2018}{Woodstock, NY}
\acmPrice{15.00}
\acmISBN{978-1-4503-XXXX-X/18/06}

\begin{document}

\title{Discreetly Exploiting Inter-session Information for Session-based Recommendation}


\author{Zihan Wang}
\affiliation{%
  \institution{Northeastern University}
  \country{China}}
\email{2101816@stu.neu.edu.cn}

\author{Gang Wu}
\affiliation{%
  \institution{Northeastern University}
  \country{China}}
\email{wugang@mail.neu.edu.cn}

\author{Haotong Wang}
\affiliation{%
  \institution{Northeastern University}
  \country{China}}
\email{2171931@stu.neu.edu.cn}





\renewcommand{\shortauthors}{Trovato and Tobin, et al.}

\begin{abstract}
Limited intra-session information is the performance bottleneck of the early \gls*{GNN} based \gls*{SBR} models.
Therefore, some \gls*{GNN} based \gls*{SBR} models have evolved to introduce additional inter-session information to facilitate the next-item prediction. 
However, we found that the introduction of inter-session information may bring interference to these models. 
The possible reasons are twofold. 
First, inter-session dependencies are not differentiated at the factor-level.
Second, measuring inter-session weight by similarity is not enough.
In this paper, we propose \gls*{DEISI} to solve the problems.
For the first problem, \gls*{DEISI} differentiates the types of inter-session dependencies at the \emph{factor level} with the help of \gls*{DRL} technology. 
For the second problem, \gls*{DEISI} introduces \emph{stability} as a new metric  for weighting inter-session dependencies together with the similarity.
Moreover, \gls*{CL} is used to improve the robustness of the model.
Extensive experiments on three datasets show the superior performance of the \gls*{DEISI} model compared with the state-of-the-art models.

\end{abstract}




\maketitle

\glsreset{DEISI}
\glsreset{DRL}
\glsreset{SBR}
\glsreset{CL}
\glsreset{GNN}

\section{Introduction} \label{Sec:Introduction} 

Nowadays recommendation are expected to meet the short-term interests of users according to specific session context \cite{devooght2017long,jannach2015adaptation}.
Therefore, the \gls*{SBR} has become a hot topic.

From a sequence processing perspective, many commonly used natural language processing (NLP) techniques are applicable to SBR, such as Recurrent Neural Network (RNN)\cite{liu2016recurrent,zaremba2014recurrent,tan2016improved}, Attention mechanism\cite{jain2019attention,galassi2020attention}, and \gls*{GNN} \cite{schlichtkrull2020interpreting,wu2023graph,lai2022attribute}.
Due to better complex relationship modeling capabilities, GNN-based SBR models have become the current mainstream. 
Early GNN-based SBR models \cite{wu2019session,yu2020tagnn,hidasi2015session,li2017neural,liu2018stamp}only consider intra-session information for the next-item recommendation. 
It limits their performance because the length of the session is usually too short for effectively learning. 
To improve the performance, some models include additional information from other sessions\cite{xia2021self,xia2021self2,zheng2019balancing,qiu2020exploiting,wang2022exploiting}. 
Such inter-session information is obtained by establishing inter-session dependencies and then used for learning as a supplement to in-session information. 

However, the introduction of inter-session information brings interference as well, which may distract the attention from the current session and mislead the recommendation.
Generally speaking, there are two kinds of problems with these GNN-based SBR models.

\begin{figure}[ht]
 \centering
 \includegraphics[scale = 0.45]{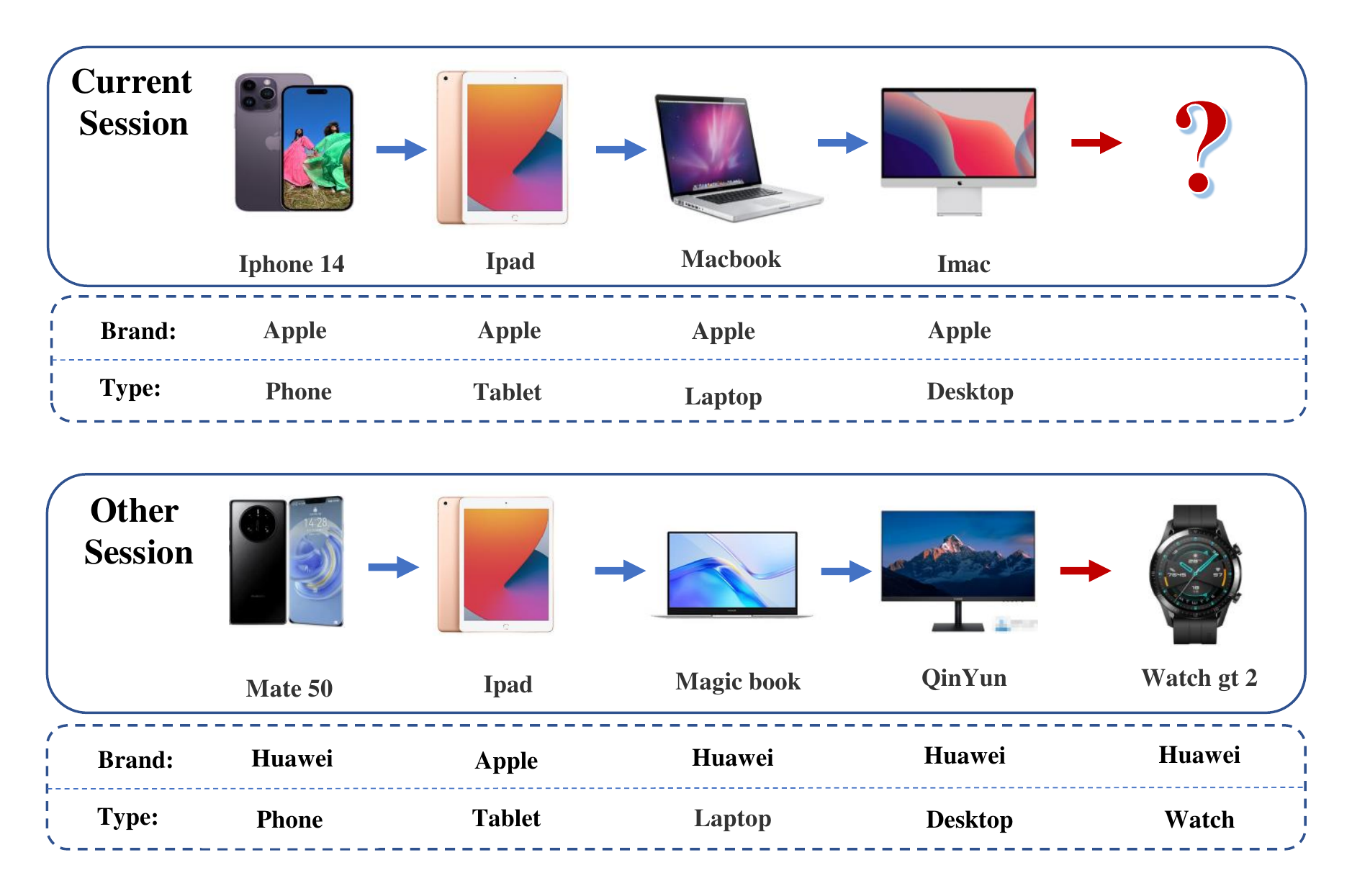}
 \caption{An illustrative example of the challenges in inter-session information learning}
 \label{fig:Example}
\end{figure}

First, inter-session dependencies are not differentiated at the factor-level. 
In other words, these models have no the ability of excluding the interference of irrelevant factors within the inter-session dependencies, but to emphasize only relevant factors.
The result is a decrease in the accuracy of inter-session dependency representation.
Take the two sessions in Figure \ref{fig:Example} for example.
Both sessions are about electronic products.
They share similar interests in type preference but distinct differences in terms of brand. 
From an overall perspective, the brand factor reduces the strength of the inter-session dependency, i.e. the similarity.
This, in turn, may lead to inadequate use of out-of-session information from the other session.

Second, the similarity is used as the only metric for weighting inter-session dependencies.
However, in some cases, the similarity is obviously not the dominant factor of GNN weight value.
For the example in Figure \ref{fig:Example}, intuitively, an appropriate next-item recommendation for the current session may be Apple Watch.
Why not recommend Huawei Watch gt2 by referring to the other session?
One main reason is that the user preferences for the current session have stabilized on Apple in terms of the brand factor.  
Therefore, refusing to learn out-of-session information about the brand factor from other sessions is a wise strategy. 

To solve the above problems, we propose \gls*{DEISI}.

For the first problem, inter-session dependencies are finely differentiated at the \emph{factor-level} in \gls*{DEISI} with the help of \gls*{DRL} technology \cite{li2022disentangled,ma2019learning,ma2020disentangled,wang2020disentangled}, which has been deployed in sequential recommendation and collaborative filter recommendation. 
\gls*{DRL} has the ability to disentangle latent factors hidden in observed data, and has been proved effective by Disen-GNN \cite{li2022disentangled} in learning independent factor-level intra-session representations, i.e. embeddings.

For the second problem, we discover and define a new metric, \emph{stability}, in \gls*{DEISI} besides the similarity.
The basic principle is that the more stable the interest of a session, the less likely it is to be influenced by other sessions where the stability is closely related to the deviation of the interests within the session.
Stability and similarity are used together to more accurately estimate the weight of inter-session dependencies. 

In addition, we adopt \gls*{CL} \cite{wei2021contrastive,xie2022contrastive,qin2021world} to improve the robustness of the model considering that data sparseness in SBR often leads to inaccurate embedding learning.

To summarize, the main contributions of this paper are as follows.
\begin{itemize}
\item \gls*{DRL} is introduced to refine the inter-session dependencies of \gls*{SBR} at the factor-level, which effectively avoids interference from irrelevant factors. 
\item A new metric, the intra-session interest stability, is defined as a supplementary of the inter-session similarity to better estimate the GNN weights for aggregating other sessions in \gls*{SBR}. 
\item 
Extensive experiments on three datasets were conducted to demonstrate the superior performance of our proposed \gls*{DEISI} model compared with the state-of-the-art models. 
\end{itemize}

\section{Related Work}\label{Sec:RelatedWork}

\subsection{Session-based Recommendation}
In recent years, Session-based Recommendation (SBR) has received more and more attention and many approaches have been proposed for this field. 
Session-based Recommendation aims to predict the user's next interaction based on only the anonymous current session. 
Early traditional models adopt Markov Decision Process(MDP) \cite{rendle2010factorizing} and their performance is not satisfying because they think that the transition of two adjacent items is very strict. 

With the vigorous development of Neural Network, many models based on it have emerged to improve the recommendation accuracy of SBR \cite{wang2020global,de2021transformers4rec,wang2019collaborative,qiu2019rethinking,sun2019bert4rec}. 
GRU4Rec applies RNN to SBR for the first time \cite{hidasi2015session}. 
It uses RNN and Gated Recurrent Unit (GRU), which can learn session representation by analyzing the sequential relationship in the session, then predict the next item most likely interacted. 
Then, NARM \cite{li2017neural} was proposed, which uses RNN and attention mechanism together. 
The attention mechanism in it is responsible for capturing the global relationship in the session because they notice that global relationship also exists in a session besides the sequential relationship. 
STAMP\cite{liu2018stamp} also adopts the attention mechanism, and it achieved better results by emphasizing more short-term interest and focusing more on users' recent clicks. 

Graph Neural Network (GNN) has achieved great success in other fields and exhibited great potential for representation learning, so it has been introduced in SBR, too. 
SR-GNN \cite{wu2019session} makes this attempt first and achieves excellent performance.
It builds a directed intra-session graph based on the sequential relationship of a session and deploys GGNN convolution on it to learn more accurate embeddings. 
TAGNN \cite{yu2020tagnn} has made further improvements on the basis of SR-GNN. 
It adds the embedding expression that is sensitive to the prediction target meaning that regarding the user's interest as a dynamically changing interest rather than a fixed one, which is more in line with the realistic scenario. 

Disen-GNN \cite{li2022disentangled} notices that previous models all ignore the fact that user interest is often driven by a certain factor(e.g. preference for the brands). 
So it introduced Disentangled Represent Learning(DRL) to learn independent factor-level embeddings and explore the factor-level intra-session information.

However, these models all face a limit, which is that the intra-session information is insufficient. 

\subsection{Utilization of inter-session information in SBR}
More scholars have noticed that rich inter-session information can be mined as a supplement to the scarce intra-session information to help us make more accurate recommendation. \cite{qiu2020exploiting,wang2022exploiting,choi2022s,ye2020cross}. 

FGNN-BCS \cite{qiu2020exploiting} first generates a global item graph to learn extra information outside the session. 
It has achieved better performance than the models that only considers intra-session information at that time. 

I3GN \cite{zheng2019balancing} connects items outside the current session and items inside by establishing an inter-session graph. 
By utilizing the inter-session graph, I3GN can learn some extra information about items in the current session. 

DHCN \cite{xia2021self} realizes that the graph contrastive learning paradigm in SBR is often ineffective because of the sparsity of the graphs. 
A novel paradigm of contrastive learning was proposed by it. 
It contains two channels, one to learn intra-session information, and the other to learn extra inter-session information, helping item learn its most essential part by comparing the information learned from the two channels.  

COTREC \cite{xia2021self2} is an improved version of DHCN and it combines contrastive learning with co-training, and a SBR enhancement framework is developed. 
The proposed contrastive graph co-training preserves the complete session information and fulfills
genuine data augmentation. 

But they fail to filter the inter-session information, which means that interfering information can also be learned by these models. 
It is with this in mind that we hope to learn only those really useful information from inter-session information and avoid the interference.

\section{Preliminaries}\label{Sec:Preliminaries}
In this section, we introduce the preliminary knowledge to help better understand of the proposed model.

\subsection{Problem Statement of \gls*{SBR}}
The basic task of recommendation systems is to predict the user's next interacted item. 
In traditional recommendation scenarios, historical interactions are always available for models to capture users' long-term interests.
However, in some cases, it is hard to obtain historical information, e.g., cold start and privacy protection.
Unlike previous recommendation scenarios, \gls*{SBR} makes predictions solely based on the current session.  

Let $V = \left\{ v_1, ..., v_{N} \right\}$ denote the set of items where $N$ is the number of items. 
An anonymous session is represented as a sequence $s = \left[ v_{(s,1)}, ..., v_{(s,n)}\right]$ ordered by timestamps. 
$v_{(s,i)}\in V$ is the $i$-th item that an anonymous user has interacted in the session $s$. 
SBR models intend to process the information contained in $s$ to construct the user's interest. 
We generate probabilities $\hat{y}$ by for all possible items based on input session $s$. 
Each element’s value of vector $\hat{y}$ is the recommendation score of the corresponding item. 
The recommendation scores are usually the similarity between the user's interest and the item embeddings. 
Finally, we select those with the highest scores as the prediction for the user's next interaction item $v_{(s,n+1)}$. 

\subsection{Disentangled Represent Learning (\gls*{DRL})}
The purpose of \gls*{DRL} is to learn factor-level representation, usually by splitting the original embedding into multiple independent dimensions, so that each dimension can represent more fine-grained semantics. 
The introduction of \gls*{DRL} can improve the interpretability and \todo{the robustness} of the model.

In SBR, the input of DRL is the embedding of either an item or a session. 
Let $\bm{c}\in \mathbb{R}^d$ represent a $d$ dimension embedding. 
Then DRL computes $k$ factor-level embeddings from $\bm{c}$ by re-embedding it into corresponding spaces. 
The factor-level embedding $f^t$ on factor $t$ is computed with Equation \ref{Equ:DRL}.

\begin{equation}\label{Equ:DRL}
    \bm{f}^t = \sigma (\bm{c}^\top \bm{W}_t) + \bm{b}_t,     (1\leq t \leq k)
\end{equation}

Here, $\bm{W}_t \in \mathbb{R}^{d \times d_f}$ and $\bm{b}_t \in \mathbb{R}^{d_f}$ are the weight matrix and bias on factor $t$. 
And $d_f = \lfloor \frac{d}{k} \rfloor$ is the dimension of factor-level embedding.

In order to avoid redundant information between factors, DRL uses the following loss function as the learning objective to generate independent factor-level embeddings. 
\begin{equation}\label{Equ:re_emb}
    \mathcal{L}_d = \sum_t^k\sum_{j\neq t}^k \bm{dCor}(f^t, f^j)
\end{equation}
$\bm{dCor}$ is a formula to measure the correlation between variables in different spaces. 
For other details, please refer to \cite{GJ2007Measuring}. 

\subsection{Inter-session and Intra-session Information}
\todo{As we mentioned earlier, SBR is proposed for an anonymous session. 
Any information contained in the session is intra-session information. 
}
\todo{Inter-session information is from other sessions. 
It can help us make recommendations as the additional information.
Take a simple example, the item $v_i$ appears in the current session. 
And other session have both $v_i$ and $v_j$. 
Hence, we can learn that $v_j$ has a certain correlation with $v_i$, so maybe recommend $v_j$
to the current session is satisfactory. }

\section{Methodology}\label{Sec:Methodology}
In this section, we present the proposed \gls*{DEISI} model. 
Necessary formulas are used to assist in stating the principle behind the processing of each module. 

\begin{figure*}[htp]
 \centering
 \includegraphics[width=0.99\textwidth]{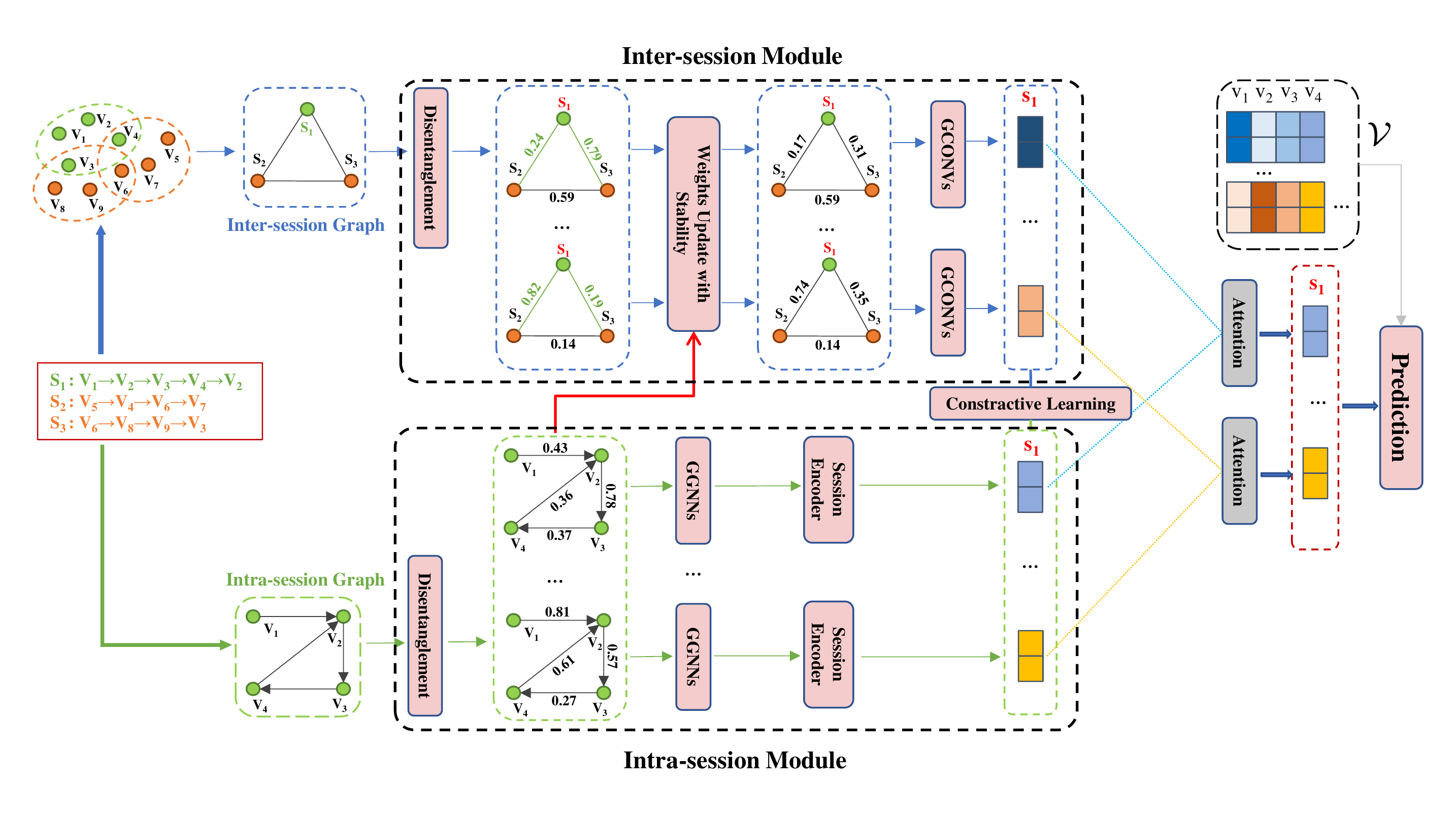}
 \caption{Overview of \gls*{DEISI}}
 \label{fig:DEISI-Models}
\end{figure*}

Figure \ref{fig:DEISI-Models} gives an overview of \gls*{DEISI}.
Besides the necessary \emph{initialization}, \gls*{DEISI} contains two key modules, namely, the \emph{intra-session} module, the \emph{inter-session} module. 
Both modules both contain $k$ factor-level channels ($k$ is the total number of disentangled factors), which are respectively responsible for embedding corresponding information of each factor. 
Then, factor-level embeddings obtained from both modules are aggregated to get the final representation of the session interest for further making the next-item prediction.

\subsection{Initialization} \label{SubSec:Initialization}
The inter-session graphs, the intra-session graphs, and the embeddings of items and sessions are constructed in the initialization module.


\subsubsection{Inter-session Graph}
Let $S = \left\{ s_1, ..., s_M\right\}$ be $M$ sessions input in a batch. 
An inter-session graph is an undirected graph $\mathcal{G}^g_S= <\mathcal{V}_S^g, \mathcal{E}_S^g >$ representing dependency relations between sessions in $S$ where the superscript $g$ stands for \emph{global}.

The vertex set $\mathcal{V}_S^g = S$ where a vertex represents a session.

The edge set $\mathcal{E}^g_S=\{<s_i, s_j>|s_i\cap s_j \neq \emptyset\}$ where $s_i, s_j\in S$. 
That means each edge $e^g_S\in \mathcal{E}^g_S$ connects two session vertexes if they share common items.
We initialize the edge weight to be $1.0$.

\subsubsection{Intra-session Graph}
We define an intra-session graph to be a directed graph $\mathcal{G}^l_s=<\mathcal{V}_s^l, \mathcal{E}_s^l>$ for representing the inner relations in a session $s$ as in SR-GNN\cite{wu2019session} where the superscript $l$ stands for \emph{local}. 

The vertex set $\mathcal{V}^l_s=\{v|\forall v\in s\}$ are items appeared in $s$. 

The edge set $\mathcal{E}^l_s=\{<v_{(s,i)},v_{(s,i+1)}>|\forall v_{(s,i)}\in s\}$ where $v_{(s,i)}$ is the $i$-th item in session $s$.
That means each edge $e^l_s\in \mathcal{E}^l_S$ starts from the previous item and points to the next item in the sequence of session $s$.  
Take $s_1=[v1,v2,v3,v4,v2]$ in Figure \ref{fig:Example} as an example.
The edge set $\mathcal{E}^l_{s_1}= [(v_1,v_2), (v_2,v_3), (v_3,v_4), (v_4,v_2)]$.

\subsubsection{Item Embedding} 
For an input session $s = \left\{ v_{(s,1)}, ..., v_{(s,n)}\right\}$, let the initial embedding of an item $v_{(s,i)}\in s$ be $\bm{c}_{(s,i)}$.
Then \gls*{DRL} is used to obtain the interest preferences for different factors hidden behind the interactions by re-embedding each item embedding $\bm{c}_{(s,i)}$ into different spaces as Equation \ref{Equ:item_re_emb_1}.
\begin{equation}\label{Equ:item_re_emb_1}
    \bm{DRL}(\bm{c}_{(s,i)}) = \mathcal{F}_{(s,i)} = \left\{ f^1_{(s,i)}, ..., f^k_{(s,i)}\right\}
\end{equation}

Here, $f^t_{(s,i)}\in \mathbb{R}^{d_f}$ is the factor-level embedding of item $v_{(s,i)}$ on factor $t$ obtained with Equation \ref{Equ:DRL}.
And $\mathcal{F}_{(s,i)}$ is the set of all factor-level embeddings of the item $v_{(s,i)}$, 




\subsubsection{Session Embedding} 
In Equation \ref{Equ:session_emb}, the initial embedding of session $s$ is defined to be the average of all embeddings of items in $s$.
\begin{equation}\label{Equ:session_emb}
    \bm{c}_s = \frac{\sum_1^n c_{(s,i)}}{n}
\end{equation}

Similarly, \gls*{DRL} is employed for $\bm{c}_s$ to get $k$ factor-level session embeddings as shown in Equation \ref{Equ:session_re_emb}. 
\begin{equation}\label{Equ:session_re_emb}
    \bm{DRL}(\bm{c}_s) = \mathcal{F}_s =\left\{ f^1_s,... ,f^k_s\right\}
\end{equation}
where $f^t_s\in \mathbb{R}^{d_f}$ is the factor-level embedding of session $s$ on factor $t$. 

\subsection{Inter-session Module}
This module is responsible for learning factor-level inter-session embeddings with GCN.
Let $S = \left\{ s_1, ..., s_M\right\}$ be $M$ sessions input in a batch. 
The input of the inter-session module include a set $\{\mathcal{F}_{s_1},...,\mathcal{F}_{s_M}\}$ containing the all initial factor-level embeddings of all sessions and the inter-session graph $\mathcal{G}^g_S$. 

\subsubsection{Factor-level Inter-session Dependency}
As mentioned in the introduction, two sessions may have different strength of dependencies on different factors.
Therefore, in order to learn factor-level embeddings of dependencies, 
a set of $k$ factor-level inter-session graphs $\left\{ \mathcal{G}^{(g,1)}_S, ..., \mathcal{G}^{(g,k)}_S \right\}$ are derived from $\mathcal{G}^g_S$. 

These factor-level inter-session graphs have exactly the same vertexes and edges.
The only difference exists in the edge weights, which are affected by the factor-wise cosine similarities between the sessions. 

Let $\mathcal{A}^{(g,t)}_S$ be the adjacency matrix of graph $\mathcal{G}^{(g,t)}_S$, and $a_{ij}^{(g,t)}$ be the weight of edge between $s_i$ and $s_j$ on factor $t$.
Then we define the weight with Equation \ref{Equ:interweight}.
\begin{equation}\label{Equ:interweight}
    a_{ij}^{(g,t)} = \frac{f^t_{s_i}f^t_{s_j}}{\lvert\lvert f^t_{s_i}\rvert \rvert\cdot \lvert\lvert f^t_{s_j}\rvert\rvert}
\end{equation}
Note that the connections established at this time is only based on similarity. 

\begin{figure}[ht]
 \centering
 \includegraphics[scale = 0.7]{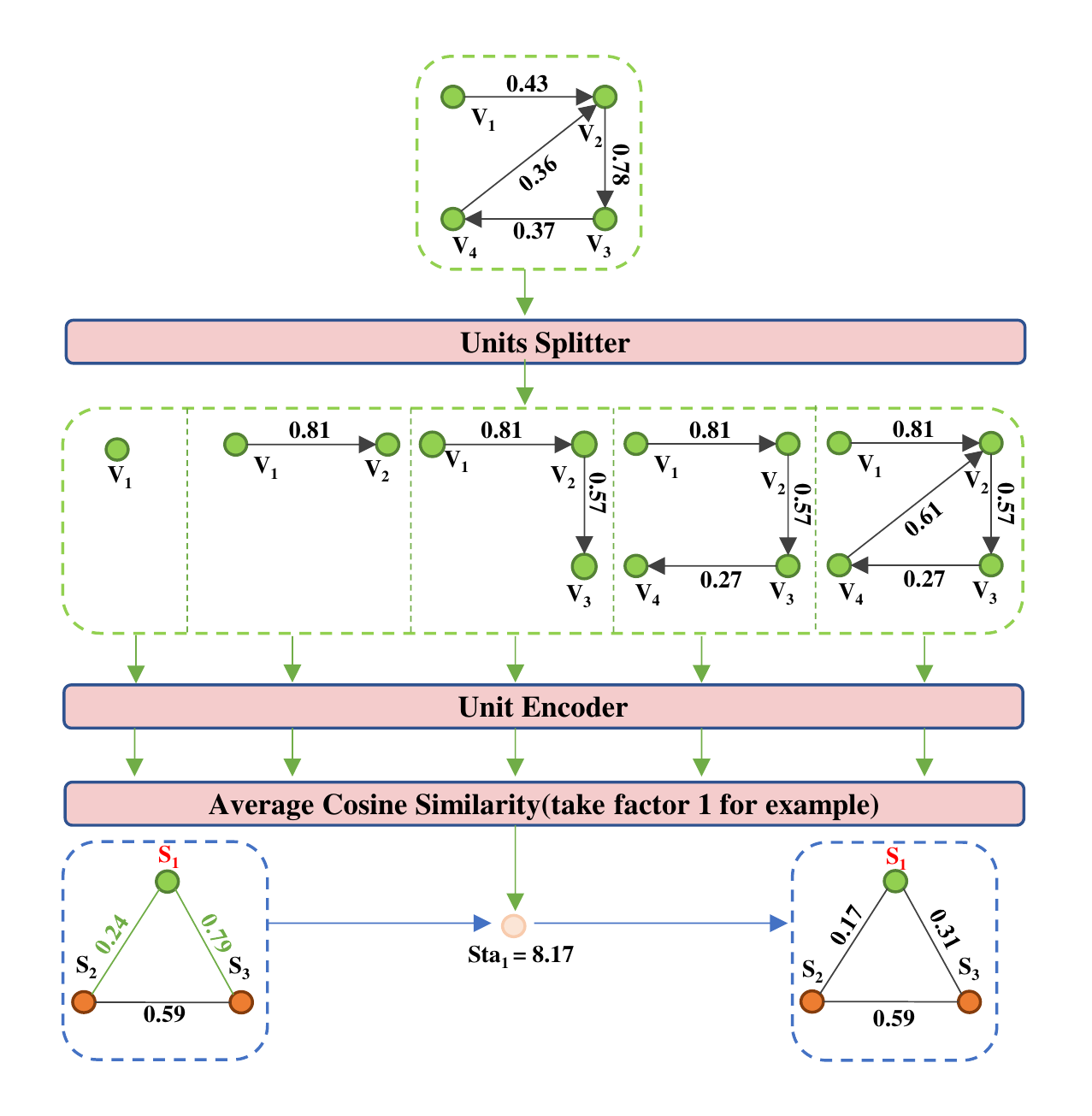}
 \caption{Details for weight update with Stability}
 \label{fig:Item}
\end{figure}

\subsubsection{Weight Update with Stability}
As mentioned before, another feature of \gls*{DEISI} is to incorporate interest stability in weighting inter-session dependencies as Figure \ref{fig:Item}. 
We propose a scheme to measure the stability according to the interest bias of the interest units.
Here, the $i$-th interest unit $U_{(s,i)}$ is defined to be a sub-sequence of session $s$ as in Equation \ref{Equ:interest-unit}.
Obviously, there are $n$ interest units split from $s$, where the $i$-th interest unit contains the first $i$ items in $s$, and the next interest unit has one more item $v_{(s,i+1)}$ than the previous interest unit. 

\begin{equation}\label{Equ:interest-unit}
    U_{(s,i)} = \left\{ v_{(s,1)},..., v_{(s,i)}\right\}
\end{equation}

To measure the stability, the factor-level embeddings of these interest units are required. 
Similarly, we calculate the factor-level embeddings using DRL based on the initial average-pooling embeddings. 
Let $\mathrm{UnitEncoder}$ be a function for this process. 
\begin{equation}
   \mathrm{UnitEncoder}(U_{(s,i)})=\left\{U^1_{(s,i)},... ,U^k_{(s,i)}\right\}
\end{equation} 

Then, the stability of session $s$ on factor $t$ can be measured with Equation \ref{Equ:Stability}, which accumulates the cosine similarities of the embeddings of all interest units as the divergence on factor $t$. 

\begin{equation}\label{Equ:Stability}
     \frac{1}{{Sta}^t_s} =  \sum_{i=1}^n\sum_{j\neq i}^n\frac{U^t_{(s,i)}U^t_{(s,j)}}{\lvert\lvert U^t_{(s,i)}\rvert\rvert \cdot \lvert\lvert U^t_{(s,j)}\rvert\rvert}
\end{equation}
\begin{equation}
     [\frac{1}{{Sta}^1_s},...,\frac{1}{{Sta}^k_s}] = \mathrm{SoftMax}([\frac{1}{{Sta}^1_s},...,\frac{1}{{Sta}^k_s}])
\end{equation}

The higher the ${Sta}^t_s$, the more stable the interest preference for factor $t$, and hence less attention should be paid on this factor. 

With the introduction of ${Sta}^t_s$, the weight of $\mathcal{A}^{(g,t)}_S$ is updated as in Equation \ref{Equ:weightupdate}. 

\begin{equation}\label{Equ:weightupdate}
     a^{(g,t)}_{ij} = \sqrt{\frac{a^{(g,t)}_{ij}}{{Sta}^t_{s_i}}}
\end{equation}
where $a^{(g,t)}_{ij}$ denotes the any weight in the $\mathcal{A}^{(g,t)}_S$. 
Due to the difference in stability, each session has its own inter-session graph with the adjacency matrix $\mathcal{A}^{(g,t)}_s$. 

\subsubsection{Convolution on factor-level inter-session graph}
Based on the established factor-level inter-session dependencies, graph propagation is employed to update the vertexes' embeddings for learning inter-session information. 
As shown in Equation \ref{Equ:convolution},

\begin{equation}\label{Equ:convolution}
    \mathcal{F}^{(g,t)^{(l+1)}}_S = Norm(\sigma(\mathcal{F}^{(g,t)^{(l)}}_S\mathcal{A}^{(g,t)}_S) + \mathcal{F}^{(g,t)^{(l)}}_S)
\end{equation}
where $\left\{ \mathcal{A}^{(g,1)}_S, ..., \mathcal{A}^{(g,k)}_S \right\}$ are the adjacency matrices of factor-level inter-session graphs, and $\mathcal{F}^{(g,t)^{(l)}}_S$ is embeddings after $l$ propagations. 
In the end, the output is the concatenation of factor-level session embeddings $\mathcal{F}*^g_s = [\mathcal{F}*^{(g,1)}_s,...,\mathcal{F}*^{(g,k)}_s]$. 

\subsection{Intra-session Module}
This module is used to learn the current session's embeddings by aggregating the intra-session information of the items that has been interacted with. 
Intuitively, each interacted item has impacts on the session embedding in various factors. 
Hence we deploy $k$ channels as in the inter-session module to process corresponding factor-level embeddings. 
The input here includes the set consisting of factor-level embeddings of all items in $s$, i.e., $\{\mathcal{F}_{(s,1)},...,\mathcal{F}_{(s,M)}\}$, and the set of factor-level intra-session graphs $\left\{\mathcal{G}^{(l,1)}_s, ..., \mathcal{G}^{(l,k)}_s \right\}$ derived from $\mathcal{G}^l_s$.

In each factor channel, we adopt the same processing method as SR-GNN for its excellent performance. 
Graph gated neural network (GGNN) is employed to update the embeddings at first, GGNN has a built-in gated mechanism similar to GRU, which can further \todo{screen} information during the propagation process. 
\todo{For details of GGNN, please refer to \cite{wu2019session}. }
The final layer's output of GGNNs are $\left\{ f*^t_{(s,1)}, ..., f*^t_{(s,n)}\right\}$. 
Then soft attention is adopted as the session encoder. 
\begin{equation}
    \alpha^t_i = q^\top \sigma(f*^t_{(s,i)}\top\bm{W}^{(1)}_t + f*^t_{(s,n)}\top W^{(2)}_t )
\end{equation}
\begin{equation}
    \mathcal{F}^{(v,t)}_{s_g} =\sum^n_{i=1}\alpha^t_i f*^t_{(s,i)}
\end{equation}
\begin{equation}
    \mathcal{F}*^{(v,t)}_s = \bm{W}^{(3)}_t[\mathcal{F}^{(l,t)}_{s_l},\mathcal{F}^{(l,t)}_{s_g}]
\end{equation}
where $\bm{q} \in \mathbb{R}^d_f $, $\bm{W}^{(1)}_t \in  \mathbb{R}^{d_f \times d_f}$, $W^{(2)}_t \in  \mathbb{R}^{d_f \times d_f}$ and $\bm{W}^{(3)}_t \in  \mathbb{R}^{d_f \times 2d_f}$ are learnable parameters. 
$\mathcal{F}^{(l,t)}_{s_l}$ and $\mathcal{F}^{(l,t)}_{s_g}$ represent the session's local and global preferences for factor $t$ respectively. 
And $\mathcal{F}^{(l,t)}_{s_l}$ is $f*^t_{(s,n)}$, the last item's factor-level embedding, such setting can make the model pay more attention to the last clicked item, because usually the last item is more related to the item that the user finally needs. 

In the end, the output here is $\mathcal{F}*^l_s = [\mathcal{F}*^{(l,1)}_s,...,\mathcal{F}*^{(l,k)}_s]$. 

\subsection{Contrastive Learning}
In order to get more accurate embeddings, we embedded a contrastive learning method. 
\todo{Contrastive Learning is a self-supervised learning method, which is used to learn the essential characteristics of objects by letting the model learn which objects are similar or different. 
Then update the embeddings by comparing the information learned from two different views. }

Here we deployed the CL strategy similar with \cite{xia2021self}.
By comparing the two embeddings of session $s$ obtained from inter-session and intra-session modules to generate self-supervision signals, the model can acquire more information for better recommendations. 
A standard binary cross-entropy (BCE) loss function has been chosen as our learning objective to measure the difference between the two. 
\begin{equation}
    \mathcal{L}_c = - \log~\sigma(H(\mathcal{F}*^g_s, {F}*^l_s) - \log~\sigma(H(\mathcal{F}*^l_s,
    \Tilde{\mathcal{F}*^l_s)})
\end{equation}
Where $\Tilde{\mathcal{F}*^l_s}$ is row-wise of column-wise shuffling type of $\mathcal{F}*^l_s$. 

\subsection{Information Combination} 
After learning information through two modules, we also need to aggregate these information. 
Inspired by \cite{zheng2019balancing}, we use fusion gating mechanism to finish this purpose .
In the end, we obtain the final session representation $\mathcal{F}*^t_s$:
\begin{equation}
     \zeta = \sigma(W^{(1)}_{(a,t)}\mathcal{F}*^{(g,t)}_s + W^{(2)}_{(a,t)}\mathcal{F}*^{(l,t)}_s)
\end{equation}
\begin{equation}
    \mathcal{F}*^t_s = \zeta \cdot (\mathcal{F}*^{(g,t)}_s) + (1-\zeta) \cdot \mathcal{F}*^{(l,t)}_s
\end{equation}
Where $W^{(1)}_{(a,t)}$ and $W^{(2)}_{(a,t)}$ are the weight matrices in fusion gate, $\zeta$ is controlling the overall proportion of information from both parties. 
The final output here is the concatenation of the outputs of $k$ channels $[\mathcal{F}*^1_s,... , \mathcal{F}*^k_s]$. 

\subsection{Prediction and Loss Function}
After the information combination, we get the final embedding of the user's preference on different factors. 
Then calculate the recommendation scores by comparing the similarity between each candidate item and the user's preference. 
Similarly, we divide this module into $k$ channels. 

The final score of a item is the sum of the scores from $k$ channels. 
The higher the score of a item, the higher the priority of the item in the recommendation. 
For the prediction loss, we use the cross-entropy as the loss function, which has been extensively used in the recommendation system: 
\begin{equation}
    \mathcal{L}_p = -\sum^N_{i=1}y_ilog(\bm{\hat{y_i}}) + (1-y_i)log(1-\bm{\hat{y}}_i)
\end{equation}
Where $\hat{y_i}$ is the final prediction of our model and $y_i$ is the one-hot vector corresponding to the real user's next interaction. 

So, the loss of the whole model consists of three parts: DRL, CL, and the prediction. 
$\beta_1$ and $\beta_2$ are controlling their proportions. 
\begin{equation}
    \mathcal{L} = \mathcal{L}_p + \beta_1 \cdot \mathcal{L}_d + \beta_2 \cdot \mathcal{L}_c
\end{equation}
Adam is adopted as the optimization algorithm to analyze the loss.

\section{Experiments}\label{Sec:Experiments}
In this section, 
we conducts several experiments to evaluate the performance of our proposed model \gls*{DEISI}.
Our experiments intend to answer the following research questions: 
\begin{itemize}
\item \textbf{RQ1}: Does \gls*{DEISI} have any performance improvement compared with baselines? Is the improvement significant?
\item \textbf{RQ2}: What is the impact of the main hyperparameters and the main components on performance. 
\item \textbf{RQ3}: How about the recommendation prediction accuracy of \gls*{DEISI} for sessions of different lengths.  
\item \textbf{RQ4}: Whether the newly proposed metric is universal and whether introducing it into other models can improve performance.
\end{itemize}

\subsection{Experiments Settings} 
In this section, we introduce some details of our experiments settings. 
\subsubsection{Baselines}
To demonstrate the comparative performance of \gls*{DEISI}, we choose several representative and/or state-of-the-art models. 
They can be categorized into three types: 
(1) \textbf{Non-GNN models}: FPMC, NARM, GRU4Rec, and STAMP; 
(2) \textbf{GNN models only considering intra-session information}: SR-GNN, Disen-GNN, and TAGNN; 
(3) \textbf{GNN models considering both inter- and intra-session information}: DHCN, I3GN, and COTREC. 

We briefly introduce them as follows. 
\begin{itemize}
\item \textbf{FPMC}\cite{rendle2010factorizing} utilizes Markov chain to make recommendations. 
It only focuses on the sequential relationship between interacted items and doesn't construct the user's latent interest. 
\item \textbf{GRU4REC}\cite{hidasi2015session} adapts GRU from NLP to SBR.
As an RNN-based model, it only cares about sequential relationships between items. 
\item \textbf{NARM}\cite{li2017neural} combined attention mechanism with Gated Recurrent Unit(GRU) to consider both global relationships and sequential relationships to make recommendations. 
\item \textbf{STAMP}\cite{liu2018stamp} emphasizes the impact of the short time and it designed a special attention mechanism with MLP. 
\item \textbf{SR-GNN}\cite{wu2019session} introduced GGNN to obtain item embeddings by information propagation and also employs a soft-attention mechanism to get the session embedding for making recommendations. 
\item \textbf{Disen-GNN}\cite{li2022disentangled} deployed DRL into SBR to learn the latent factor-level embeddings. Then use the GGNN in each factor channel to learn factor-level session embeddings. Its model structure is very simple, but it has become a state-of-the-art model. 
\item \textbf{TAGNN}\cite{yu2020tagnn} is an improved variant of SR-GNN. It replaced the original session encoder of SR-GNN with a target-aware attentive network. 
\item \textbf{DHCN}\cite{xia2021self} note that the traditional CL strategy has limited effect in SBR due to the extreme data sparsity. Therefore, it proposed a new type of CL network, which realizes data augmentation by learning inter-session information. 
\item \textbf{I3GN}\cite{zheng2019balancing} simultaneously learn inter-session and intra-session information. Synthesize the information learned from them through an attention layer.
\item \textbf{COTREC}\cite{xia2021self2} is an improved variant of DHCN, it integrated the idea of co-training into CL by adding divergence constraints to DHCN's CL module to comprehensively learn inter and intra-session information. 
\end{itemize}

\subsubsection{Datasets} 
To verify the effectiveness of \gls*{DEISI}, we conducted experiments on three commonly used datasets in a session-based recommendation system, \emph{Nowplaying}\footnote{http://dbis-nowplaying.uibk.ac.at/\#nowplaying}, \emph{Yoochoose 1/64}\footnote{http://2015.recsyschallenge.com/challege.html}, and \emph{Diginetica}\footnote{http://cikm2016.cs.iupui.edu/cikm-cup}. 
Nowplaying is a music datasets including users' listening behaviors. 
Yoochoose is released by the Recsys Challenge containing users' clicking behaviors on the e-commerce website Yoochoose. 
Diginetica contains users' transaction data and is released by CIKM Cup 2016. 2. 

We follow the commonly adopted procedures as our baselines\cite{wang2020global,wu2019session} to preprocess the datasets. 
Specifically, we removed the sessions with only one item and infrequent items that appear less than 5 times in each dataset. 
Similar to previous work, we divided training data and test data by time.  
The sessions of the last week are used as the test data for Diginetica and Nowplaying datasets. 
And the sessions of the last day are used as the test data for the Yoochoose dataset.
Notice that, only the most recent 1/64 data of Yoochoose is used because of its large size. 
So the remaining data of them is used as the training data. 
The statistics of the three datasets are exhibited in Table \ref{datasets}. 

\begin{table}[]
\centering
\resizebox{0.47\textwidth}{!}{%
\begin{tabular}{cclclcl}
\hline
\textbf{Statistics} & \multicolumn{2}{c}{\textbf{Nowplaying}} & \multicolumn{2}{c}{\textbf{Yoochoose1/64}} & \multicolumn{2}{c}{\textbf{Diginetica}} \\ \hline
\textbf{\#interactions} & \multicolumn{2}{c}{1,367,963} & \multicolumn{2}{c}{557,248} & \multicolumn{2}{c}{982,961} \\
\textbf{\#training sess} & \multicolumn{2}{c}{825,304} & \multicolumn{2}{c}{369,859} & \multicolumn{2}{c}{719,470} \\
\textbf{\#test sess} & \multicolumn{2}{c}{89,824} & \multicolumn{2}{c}{55,898} & \multicolumn{2}{c}{60,858} \\
\textbf{\#items} & \multicolumn{2}{c}{60,417} & \multicolumn{2}{c}{16,766} & \multicolumn{2}{c}{43,097} \\
\textbf{avg. length} & \multicolumn{2}{c}{7.42} & \multicolumn{2}{c}{6.16} & \multicolumn{2}{c}{5.12} \\ \hline
\end{tabular}%
}
\caption{Statistical results of datasets}
\label{datasets}
\end{table}
\subsubsection{Evaluation Metrics}
Following our baselines, we chose widely used ranking metrics P@$K$(Precise) and M@$K$(Mean Reciprocal Rank) to evaluate the recommendation results where $K$ is 10 or 20. 

\subsubsection{Hyperparameters}
According to the dimension and batch settings of the baselines, the embedding dimension and the batchsize are both set to 100. 
The learning rate is set to 0.001 which is the same as Disen-GNN. 
The number of factors $k$ is set to 5 for Diginetica dataset and Yoochoose1/64 dataset, and to 10 for Nowplaying dataset. 

\subsection{Overall Performance(RQ1)}
Table \ref{Tab:Experiments-Overall} shows the overall performance of \gls*{DEISI} compared to the baseline models. 
We take the average of 10 runs as the result. 

\begin{table*}[htbp]
\centering
\resizebox{\textwidth}{!}{%
\begin{tabular}{c|cccc|cccc|cccc}
\hline
\multirow{2}{*}{\textbf{Method}} & \multicolumn{4}{c|}{\textbf{Nowplaying}} & \multicolumn{4}{c|}{\textbf{Yoochoose1/64}} & \multicolumn{4}{c}{\textbf{Diginetica}} \\
 & P@10 & M@10 & P@20 & M@20 & P@10 & M@10 & P@20 & M@20 & P@10 & M@10 & P@20 & M@20 \\ \hline
FPMC & 0.0601 & 0.0259 & 0.0736 & 0.0282 & 0.3015 & 0.0972 & 0,4562 & 0.1501 & 0.1539 & 0.0619 & 0.2639 & 0.0689 \\
NARM & 0.1725 & 0.0601 & 0.1854 & 0.0693 & 0.5920 & 0.2495 & 0.6811 & 0.2855 & 0.3544 & 0.1513 & 0.4970 & 0.1618 \\
GRU4Rec & 0.0632 & 0.0415 & 0.0792 & 0.0449 & 0.5011 & 0.1789 & 0.6063 & 0.2288 & 0.1789 & 0.0730 & 0.2939 & 0.0829 \\
STAMP & 0.1590 & 0.0531 & 0.1766 & 0.0688 & 0.6190 & 0.2583 & 0.6874 & 0.2967 & 0.3291 & 0.1378 & 0.4539 & 0.1429 \\ \hline
SR-GNN & 0.1533 & 0.0607 & 0.1776 & 0.0749 & 0.6197 & 0.2651 & 0.7055 & 0.3094 & 0.3669 & 0.1538 & 0.5059 & 0.1750 \\
Disen-GNN & {\ul 0.1821} & 0.0759 & {\ul 0.2212} & 0.0819 & 0.6236 & 0.2701 & {\ul 0.7141} & 0.3120 & 0.3981 & 0.1769 & 0.5341 & 0.1879 \\ 
TAGNN & 0.1731 & 0.0699 & 0.1902 & 0.0782 & 0.6231 & 0.2698 & 0.7110 & 0.3101 & 0.3803 & 0.1611 & 0.5153 & 0.1790 \\ \hline
DHCN & 0.1499 & 0.0563 & 0.1622 & 0.0590 & 0.6354 & 0.2635 & 0.7078 & 0.3029 & 0.3987 & 0.1753 & 0.5318 & 0.1844 \\
I3GN & 0.1810 & {\ul 0.0761} & 0.2111 & {\ul 0.0833} & 0.6131 & 0.2701 & 0.7137 & {\ul 0.3129} & 0.3895 & 0.1721 & 0.5225 & 0.1829 \\
COTREC & 0.1571 & 0.0581 & 0.1703 & 0.0635 & {\ul 0.6242} & {\ul 0.2711} & 0.7113 & 0.3111 & {\ul 0.4179} & {\ul 0.1812} & {\ul 0.5411} & {\ul 0.1902} \\ \hline
DEISI-GNN & \textbf{0.1932} & \textbf{0.0892} & \textbf{0.2349} & \textbf{0.0943} & \textbf{0.6732} & \textbf{0.2987} & \textbf{0.7563} & \textbf{0.3387} & \textbf{0.4392} & \textbf{0.1849} & \textbf{0.5573} & \textbf{0.1946} \\ \hline
\emph{Improv} & 6.1\% & 17.2\% & 6.2\% & 13.2\% & 7.9\% & 10.2\% & 6.3\% & 8.2\% & 5.1\% & 2.0\% & 3.0\% & 2.3\%\\
\hline
\end{tabular}%
}
\caption{Comparing the prediction performance of \gls*{DEISI} with the baselines. 
The best results in them are highlighted in bold, and the second-best results are underlined. 
We add the row \emph{Improv} to show the percentage performance improvement of our model compared to the best baseline.}
\label{Tab:Experiments-Overall}
\end{table*}

By comparing the experimental results, we can make the following four observations: 

(1) Compared with RNN-based and Attention-based models, GNN-based models obviously perform better. 
It exhibits the great capability of GNN in learning more accurate embeddings and modeling session data. 

(2) Models that additionally consider inter-session information should have had a distinct advantage over those that only consider intra-session information because of more available information. 
However, this is not the case from the experimental results. 
On Nowplaying, DHCN and COTREC both perform badly compared with SR-GNN, Disen-GNN and TAGNN. 
On Yoochoose1/64, Disen-GNN performs better than DHCN. 
On Dinetica, Disen-GNN performs better than DHCN and I3GN. 
This indicates that if not discreetly exploiting the inter-session information, it may even turn out to be interference information. 

(3) Disen-GNN performs best in models that only consider the intra-session information, and even better than some models that consider both inter-session and intra-session information.
The reason is that the introduction of the \gls*{DRL} technique helps it dive deep to explore limited intra-session dependency semantics. 

(4) \gls*{DEISI} outperforms all the baseline models in all datasets.  
Compared with the previous models with best performance, \gls*{DEISI} achieved about 10\% improvement in recommendation accuracy on the Nowplaying, about 8\% on the Yoochoose 1/64 and about 3\% on the Diginetica. 
Especially in Nowplaying, there is obvious performance improvement.
We speculate that there are three reasons for out model's improvement. 
First, \gls*{DEISI} adopts GNN to build the relationship of items and of sessions for better learning embeddings. 
Second, \gls*{DEISI} and Disen-GNN both construct dependency at the factor-level, and \gls*{DEISI} perform better because it uses extra inter-session information. 
Finally, \gls*{DEISI} takes a discreet strategy to effectively learn inter-session information while avoiding being interfered with noise and redundancy included in them. 
So it can perform best compared with those baselines.

\subsection{Impact of Hyperparameters(RQ2)}
\begin{figure}[]
\centering
\begin{subfigure}{1\linewidth}
\centering
\includegraphics[width=1\linewidth]{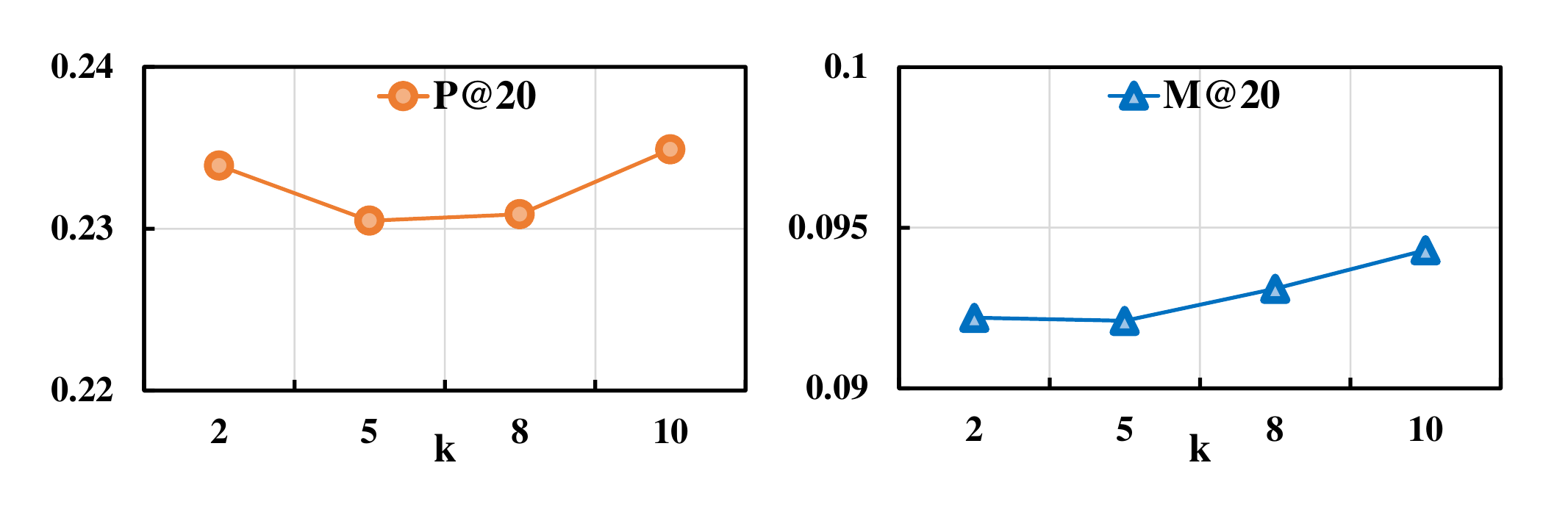}
\caption{P@20 and M@20 on Nowplaying}
\label{HYPER_NP}
\end{subfigure}

\begin{subfigure}{1\linewidth}
\centering
\includegraphics[width=1\linewidth]{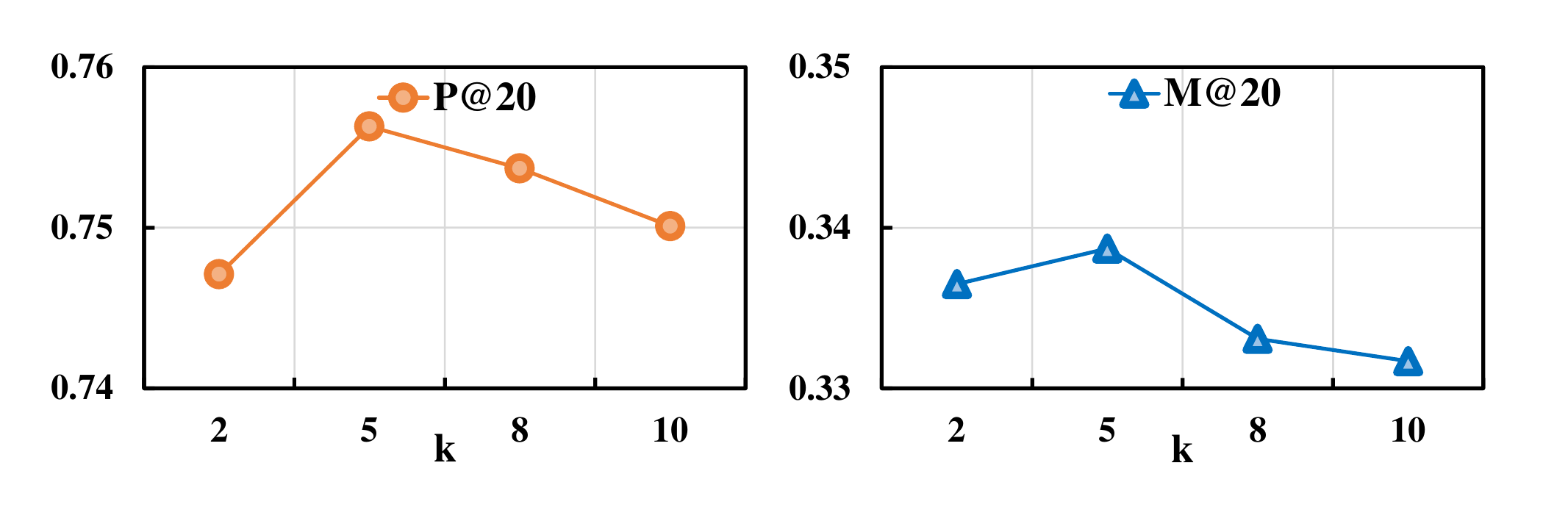}
\caption{P@20 and M@20 on Yoochoose1/64}
\label{HYPER_YC}
\end{subfigure}

\begin{subfigure}{1\linewidth}
\centering
\includegraphics[width=1\linewidth]{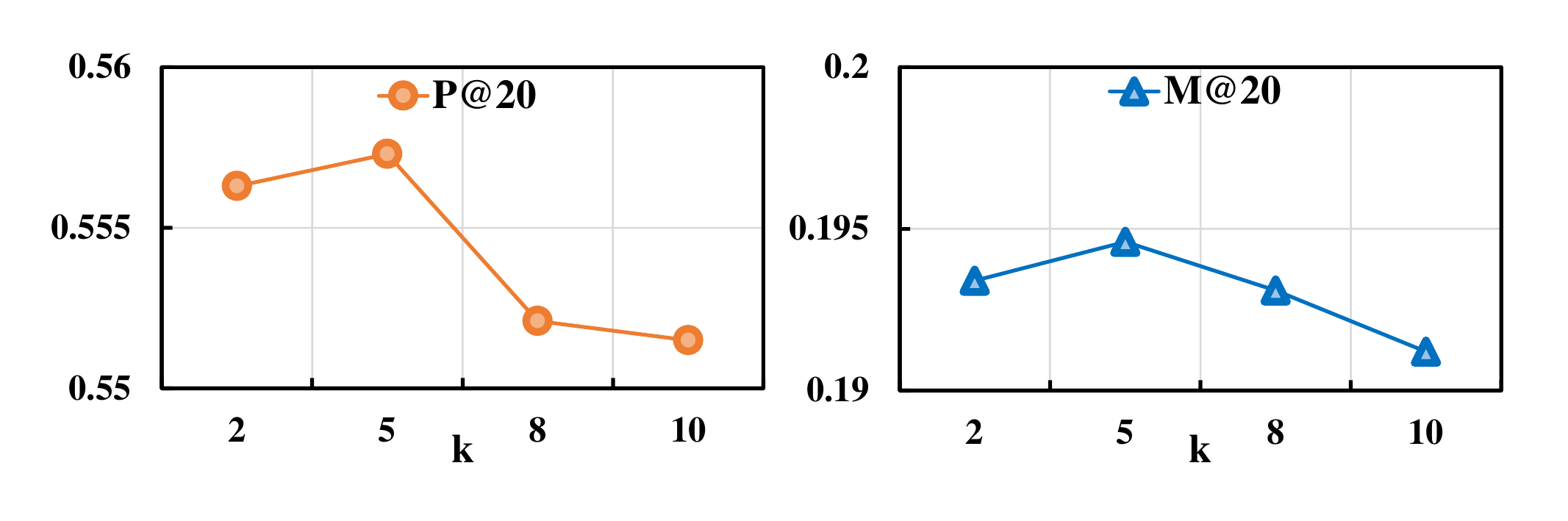}
\caption{P@20 and M@20 on Diginetica}
\label{HYPER_DG}
\end{subfigure}
\caption{Impact of the number of factors ($k$)}
\label{fig:HYPER}
\end{figure}

In the settings of experiments, the most influential hyperparameters of the model are the number of factors $k$, which is used to adjust the degree of refinement of the inter-session dependency. 
A smaller $k$ may prevent the complete separation of the inter-session dependency types.
On the other hand, a larger $k$ may cause an excessive refinement of the dependency types. 
As mentioned that, $k $ a larger $k$ brings a small factor-level embedding dimension, which may can not fully represent the semantics of the factor-level. 
To examine the impacts of different $k$ values on the performance of \gls*{DEISI}, we set $k$ to be 2, 5, 8, and 10 respectively. 
Figure \ref{HYPER_NP}-\ref{HYPER_DG} shows the corresponding model performance in the three datasets. 
In the end, we chose to set $k$ to 5 on the Diginetica dataset and Yoochoose1/64 dataset, and 10 on the Nowplaying dataset. 

\subsection{Ablation Experiments(RQ2)}
The factor-level inter-session dependency learning and the stability metric are the two key improvements of the \gls*{DEISI} model. 
In order to verify the effectiveness of them, we have designed two variants of \gls*{DEISI}.
The first one is \gls*{DEISI}~-factor, in which we remove the factor-level inter-session dependency learning from \gls*{DEISI}, but maintain the inter-session dependency from a holistic perspective. 
The second one is \gls*{DEISI}~-stability, where we set the weights between sessions solely based on their similarity. 
The rest parts of the variants keep consistent with the original \gls*{DEISI} to ensure the fairness of the ablation experiments. 

Table \ref{Tab:Ablation} shows the results.
Obviously, the two variants both have worse performance than \gls*{DEISI}. 
In other words, the proposed two key improvements are effective.
And the factor-level inter-session dependency learning has more effect on performance than the stability metric. 
The former has almost twice the impact of the latter. 

\begin{table}[]
\centering
\resizebox{0.48\textwidth}{!}{%
\begin{tabular}{c|cc|cc|cc}
\hline
\multirow{2}{*}{\textbf{Method}} & \multicolumn{2}{c|}{\textbf{Nowplaying}} & \multicolumn{2}{c|}{\textbf{Yoochoose1/64}} & \multicolumn{2}{c}{\textbf{Diginetica}} \\ \cline{2-7} 
 & P@20 & M@20 & P@20 & M@20 & P@20 & M@20 \\ \hline
DEISI & 0.2349 & 0.0943 & 0.7563 & 0.3387 & 0.5573 & 0.1946 \\ \cline{1-1}
DEISI-f & \begin{tabular}[c]{@{}c@{}}0.2238\\ (-4.7\%)\end{tabular} & \begin{tabular}[c]{@{}c@{}}0.0885\\ (-6.2\%)\end{tabular} & \begin{tabular}[c]{@{}c@{}}0.7265\\ (-3.9\%)\end{tabular} & \begin{tabular}[c]{@{}c@{}}0.3157\\ (-6.8\%)\end{tabular} & \begin{tabular}[c]{@{}c@{}}0.5371\\ (-3.6\%)\end{tabular} & \begin{tabular}[c]{@{}c@{}}0.1883\\ (-3.2\%)\end{tabular} \\ \cline{1-1}
DEISI-s & \multicolumn{1}{l}{\begin{tabular}[c]{@{}l@{}}0.2296\\ (-2.3\%)\end{tabular}} & \multicolumn{1}{l|}{\begin{tabular}[c]{@{}l@{}}0.0912\\ (-3.3\%)\end{tabular}} & \multicolumn{1}{l}{\begin{tabular}[c]{@{}l@{}}0.7410\\ (-2.0\%)\end{tabular}} & \begin{tabular}[c]{@{}c@{}}0.3277\\ (-3.2\%)\end{tabular} & \multicolumn{1}{l}{\begin{tabular}[c]{@{}l@{}}0.5498\\ (-1.3\%)\end{tabular}} & \multicolumn{1}{l}{\begin{tabular}[c]{@{}l@{}}0.1923\\ (-1.2\%)\end{tabular}} \\ \hline
\end{tabular}%
}
\caption{Comparing the prediction performance of \gls*{DEISI} with its two variants. 
The percentage of the performance degradation is put in parentheses. 
DEISI stands for \gls*{DEISI}.
DEISI-f stands for \gls*{DEISI}-factor.
DEISI-s stands for \gls*{DEISI}-stability.
}
\label{Tab:Ablation}
\end{table}

\subsection{In-depth Analysis(RQ3 and RQ4)}
\subsubsection{Performance for Different Session Lengths(RQ3)}
We studied the relation between the performance and session lengths and tried to explore whether \gls*{DEISI}'s factor-level inter-session dependency learning can somewhat solve the problems. 
Here, we performed experiments on Yoochoose1/64 dataset and Nowplaying dataset. 
We first split the test sets into \emph{long} sessions and \emph{short} sessions. 
Similar to \cite{2020Star,wu2019session}, sessions with a length $\ge$ 5 are defined to be \emph{long} sessions while the others are \emph{short} ones. 
After that, we get two sets of long and short sessions and test the performance of \gls*{DEISI} and two state-of-the-art baselines, \emph{I3GN} and \emph{COTREC}. 
After that, we compare the performance of them. 
The experimental results show that \gls*{DEISI} is superior to the other two models in both long and short sessions. 
This shows that our model comprehensively improves the accuracy of recommendation.

\begin{figure}[]
\centering
\begin{subfigure}{1\linewidth}
\centering
\includegraphics[width=1\linewidth]{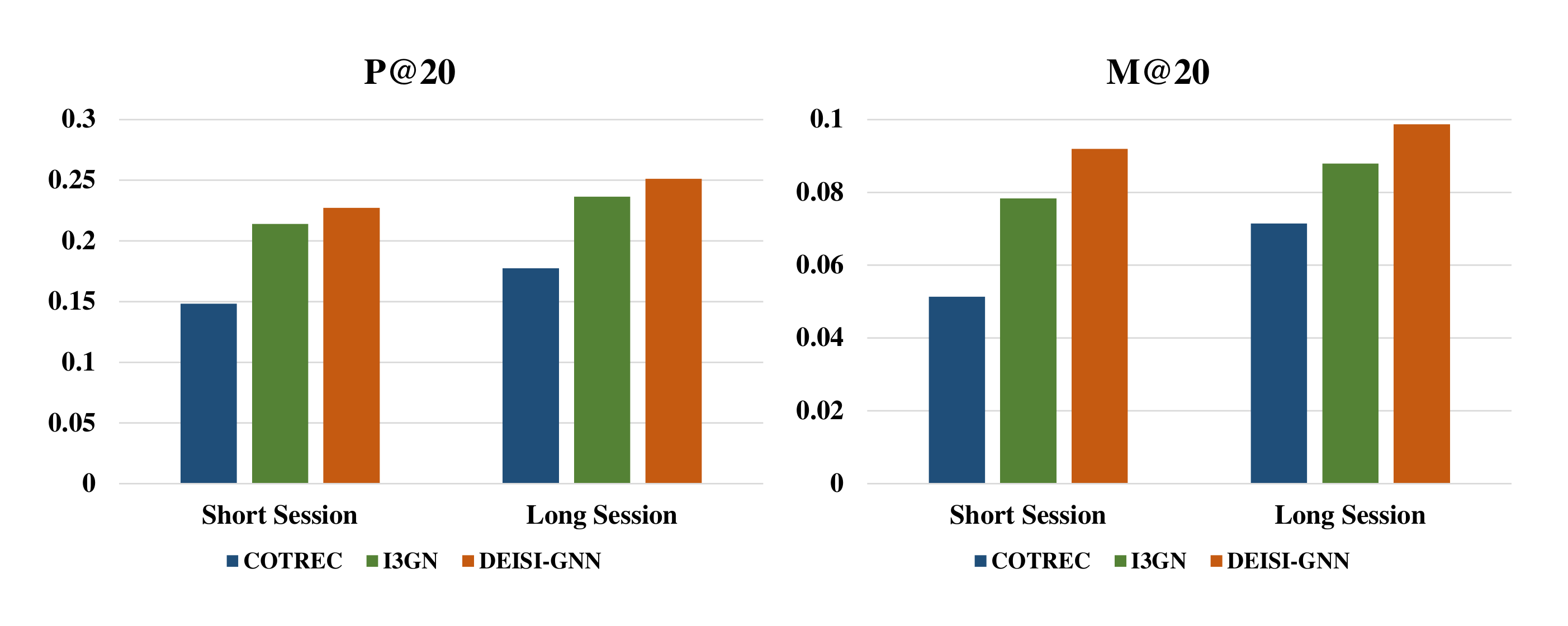}
\caption{P@20 and M@20 on Nowplaying of long and short sessions}
\label{LENGTH_NP}
\end{subfigure}

\begin{subfigure}{1\linewidth}
\centering
\includegraphics[width=1\linewidth]{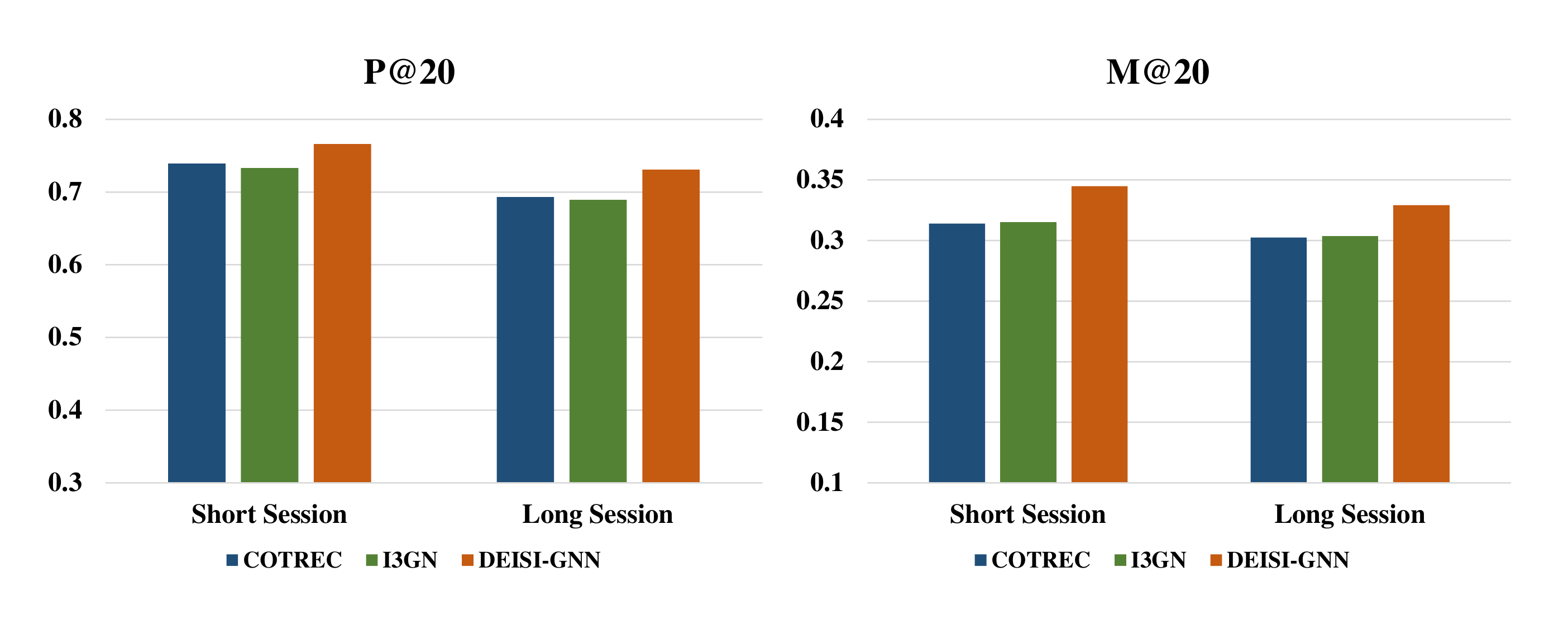}
\caption{P@20 and M@20 on Yoochoose1/64 of long and short sessions}
\label{LENGTH_YC}
\end{subfigure}

\caption{Comparison on different lengths of sessions}
\label{LENGTH}
\end{figure}

\subsubsection{Performance of Other Models with Stability(RQ4)}
To verify the effectiveness of the proposed new metric, stability, we adapted it to two other models, i.e., \emph{I3GN} and \emph{COTREC}, which also utilize additional inter-session information.

In I3GN, the strategy of using inter-session information is to create new edges between items in the current session and items in the other sessions, that is, to extract inter-session information at the item level. 
Therefore, we derive the \emph{I3GN-s} model by replacing the original I3GN weights of those newly created edges with our stability enhanced weights. 
And we conducted the experiment on Nowplaying dataset, in which the original I3GN model worked best.

For COTREC, like \gls*{DEISI}, inter-session information is extracted at the session level.
Therefore, we directly transplant our weight calculation method to COTREC to get the \emph{COTREC-s} model.
And for the similar reason, we conducted the experiment on Diginetica dataset. 

\begin{figure}[]
\centering
\begin{subfigure}{1\linewidth}
\centering
\includegraphics[width=1\linewidth]{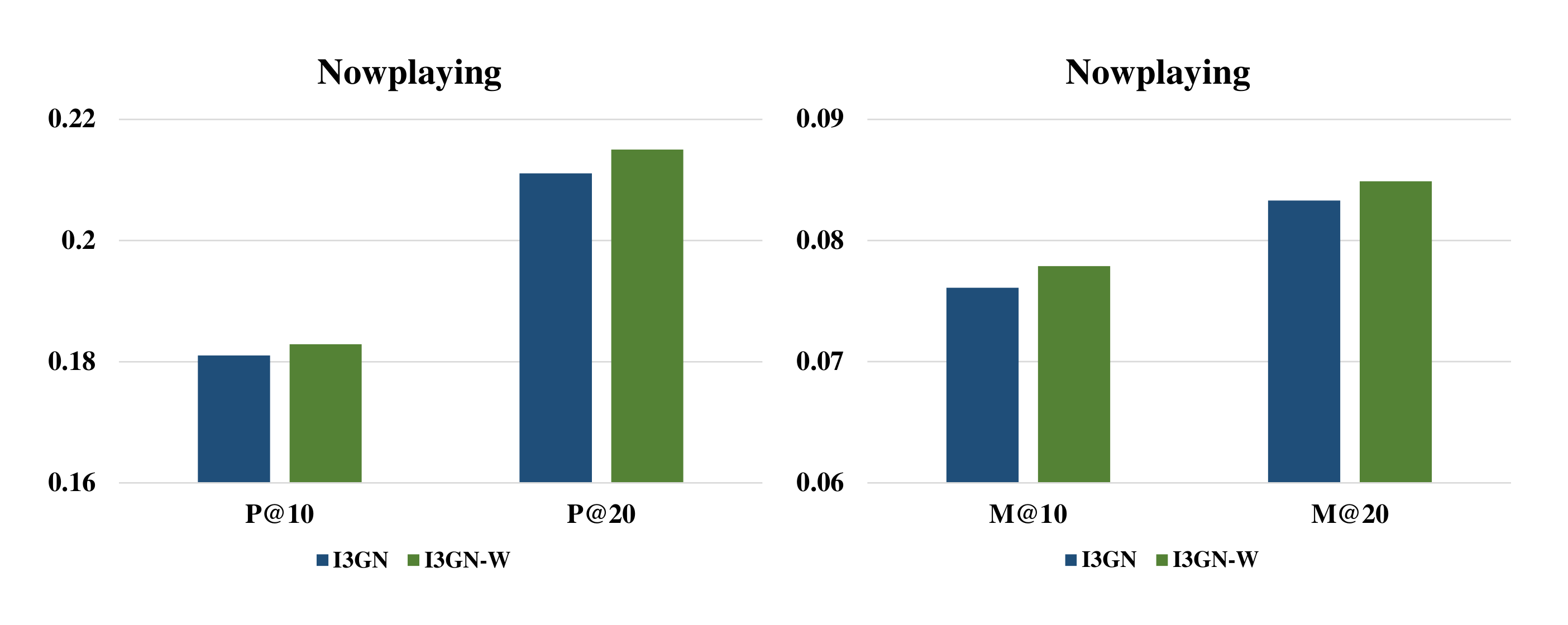}
\caption{Performance of I3GN and I3GN-s on Nowplaying}
\label{I3GN-s}
\end{subfigure}

\begin{subfigure}{1\linewidth}
\centering
\includegraphics[width=1\linewidth]{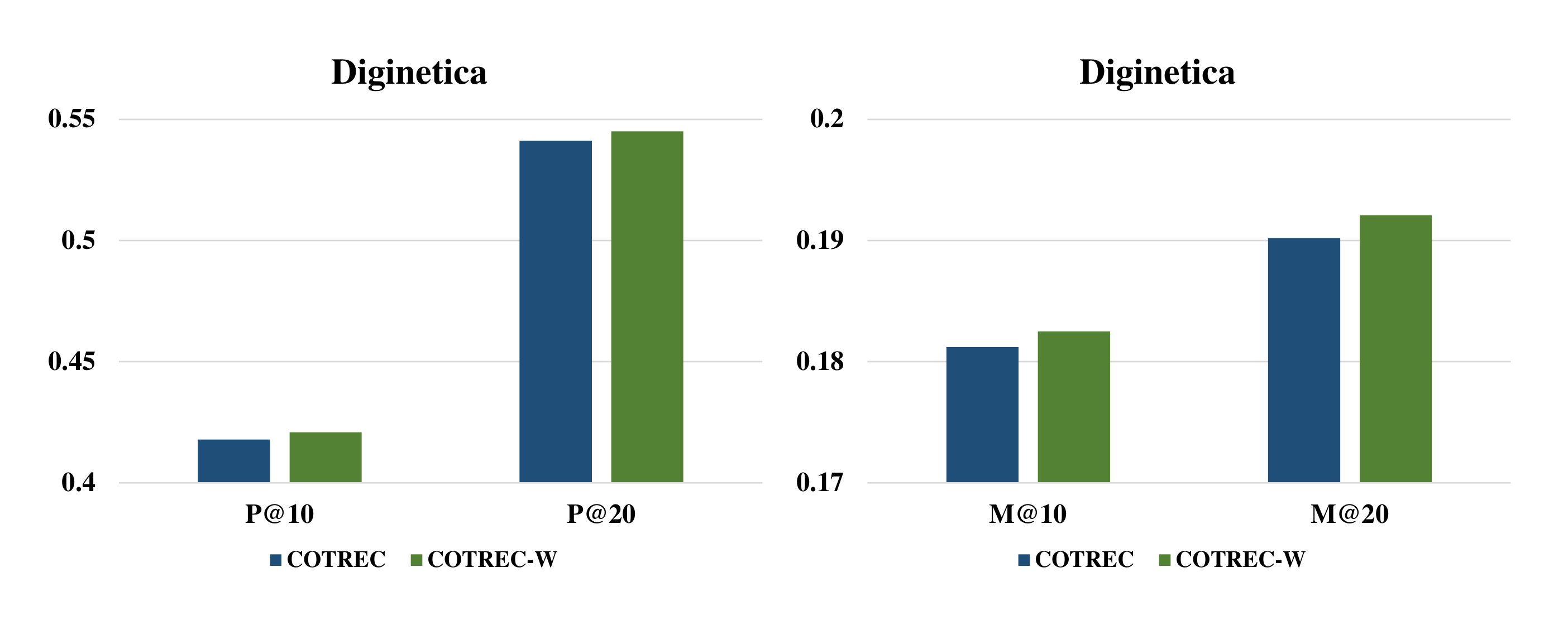}
\caption{Performance of COTREC and COTREC-s on Diginetica}
\label{COTREC-s}
\end{subfigure}

\caption{Comparison of I3GN, COTREC and their variants}
\label{Fig:OtherStability}
\end{figure}

As we can see from Figure \ref{Fig:OtherStability}, there is a slight improvement in performance when stability is taken into account in calculating the inter-session dependency weight, which illustrates the broad applicability of the new metric. 
However, the improvement is not obvious. 
The reason is that there is no factor-level inter-session dependency established for effectively discriminating between factors of different stability.

\section{CONCLUSION}\label{Sec:Conclusion} 
Nowadays, many SBR models try to use rich inter-session information to assist the recommendation. 
However, inter-session information contains both effective information and interference information. 
And these models fail to filter out these interference information to effectively help recommendations. 

Our model \gls*{DEISI} is proposed to solve these problems
\gls*{DEISI} adopts a discreet strategy to take advantage of the inter-session information and eliminate distractions while learning as much as possible. 
Specifically, \gls*{DEISI} proposes to refine the inter-session dependencies to factor-level and to consider the interest stability to adjust the weights of dependencies.

Extensive experiments were designed to prove the effectiveness of \gls*{DEISI}.
And the experimental results show that \gls*{DEISI} performs better than the state-of-the-art models and the improvements made by \gls*{DEISI} all contribute to the rise of recommendation performance. 


\end{document}